\documentclass[pra,10pt,twocolumn,showpacs,showkeys,amsmath,amssymb,superscriptaddress]{revtex4}

\usepackage{multirow} 
\usepackage{rotating}
\usepackage{mathrsfs}
\usepackage[english]{babel}  
\usepackage{amsfonts}

{\catcode`\|=\active 
  \gdef\Braket#1{\begingroup
\mathcode`\|32768\let|\BraVert\left<{#1}\right>\endgroup}}
\def\BraVert{\egroup\,\mid\,\bgroup}

\newcommand{\h}{\hspace{.1cm}}

\newcommand\T{\rule{0pt}{2.6ex}}
\newcommand\B{\rule[-1.4ex]{0pt}{0pt}}

\usepackage{cancel}
\usepackage{color}
\definecolor{Blue}{rgb}{0,0,1}
\definecolor{Red}{rgb}{1,0,0}
\definecolor{Green}{rgb}{0,1,0}
\definecolor{Purp}{rgb}{.2,0,.2}
\definecolor{white}{rgb}{1,1,1}

\begin{document}

\title{Linear Assignment Maps for Correlated System-Environment States}

\author{C\'{e}sar A. Rodr\'{i}guez-Rosario} 
\email{rodriguez@chemistry.harvard.edu}
 \affiliation{Department of Chemistry and Chemical Biology,
Harvard University, Cambridge MA, USA}
\author{Kavan Modi}
\email{kavmodi@gmail.com}
\affiliation{Centre for Quantum Technologies, National University of Singapore, Singapore}
\author{Al\'{a}n Aspuru-Guzik} 
\email{aspuru@chemistry.harvard.edu}
 \affiliation{Department of Chemistry and Chemical Biology,
Harvard University, Cambridge MA, USA}
\date{\today}
\begin{abstract}
An assignment map is a mathematical operator that describes initial system-environment states for open quantum systems. We reexamine the notion of assignments, introduced by Pechukas, and show the conditions assignments can account for correlations between the system and the environment, concluding that assignment maps can be made linear at the expense of positivity or consistency is more reasonable. We study the role of other conditions, such as consistency and positivity of the map, and show the effects of relaxing these. Finally, we establish a connection between the violation of positivity of linear assignments and the no-broadcasting theorem.
\end{abstract}
\maketitle
\section{Introduction}
The open dynamics of a quantum system is fully described by the dynamical map formalism of Sudarshan, Mathews, and Rau \cite{PhysRev.121.920, jordan:772}. The debate on the positivity \cite{choi72a,choi75} of dynamical maps began almost three decades ago \cite{Simmons80a, raggio, simmonsparkreply} and is still passionate. A significant development in this debate came due to the exchanges between Pechukas and Alicki \cite{PhysRevLett.73.1060, PhysRevLett.75.3020, PhysRevLett.75.3021}, which inspired many recent investigations into the relationships between the initial correlations of a system with its environment ($\mathsf{SE}$) \cite{PhysRevA.64.062106,jordan:052110,Shaji05a, jordan:012106,ziman06,Rodriguez07a,Kuah07a,Carteret08a,Rodriguez08,shabanilidar09a,shabanilidar09b,Modi09b}. The non-positivity of linear dynamics was given a physical interpretation in a series of papers by Jordan, Shaji and Sudarshan \cite{jordan:052110,Shaji05a, jordan:012106,  shaji05dis}. Rodr\'iguez-Rosario \emph{et al.} \cite{Rodriguez07a} showed that initially classically correlated $\mathsf{SE}$ states (as measured by quantum discord \cite{henderson01a,PhysRevLett.88.017901}) always lead to completely positive dynamics, and the converse was proved by Shabani and Lidar  \cite{shabanilidar09a}.  Without a clear mathematically interpretation of initial correlations, experimental characterization of open quantum systems dynamics is incomplete.  This debate between linearity and positivity of quantum dynamics is fundamental to our interpretation of quantum experiments \cite{Rodriguez08,rodriguezdis,Modi09a,Modidis}, the robustness of a quantum computer under decoherence \cite{zurekdecoherence}, linear error correction \cite{shabanilidar09b}, and the power of quantum computers \cite{Abrams98a}.  At the core of our understanding of quantum dynamics lies the question: what is the mathematical structure of the dynamical equations for an open quantum system?

Open quantum dynamics is the result of the reduced unitary dynamics of the system ($\mathsf{S}$) and its environment ($\mathsf{E}$). In \cite{PhysRevLett.73.1060}, Pechukas  proposed the concept of an assignment map, as a way to study the properties of the dynamical map by splitting it into the composition of three maps, $\mathcal{M}= \mathcal{T}_{\mathsf{E}}\circ \mathcal{U} \circ \mathcal{A}$.  Here, $\mathcal{M}$ is the dynamical map, $\mathcal{T}_\mathsf{E}$ is the trace with respect to the environment, $\mathcal{U}$ is the unitary map on the combined space $\mathsf{SE}$, and $\mathcal{A}$ is the assignment map, which takes a state from space $\mathsf{S}$ to the space $\mathsf{SE}$. Since both  $\mathcal{U}$ and $\mathcal{T}_\mathsf{E}$ are linear and completely positive maps, the linearity and the positivity of the dynamical map are entirely dependent on the properties of the assignment map. One of Pechukas' goals was that of showing that if the assignment is required to be linear, consistent, \emph{and} positive, then all states of $\mathsf{S}$ are mapped to a single state of $\mathsf{E}$ and the two are uncorrelated \footnote{Pechukas proof was only for two-level systems and was extended for arbitrary level systems by Jordan \cite{jordan:052110}.}, $\eta\rightarrow \mathcal{A}[\eta] =\eta\otimes T$.
\begin{table}
\begin{tabular}{c|cccc}
&	& \multicolumn{3}{c}{\bf Assignment \T\B}\\\cline{1-5}
\cline{1-5}
\multirow{4}{5 mm}{\h 
\begin{sideways}\parbox{20 mm}{\bf Correlations}\end{sideways}
	\h}&
&\h {\sl Linear} \T\B \h & \h {\sl Consistent} \h 
& \h {\sl Positive} \h \\ \cline{3-5}
& {\sl None}	\T\B    
&	\multicolumn{1}{|c|}{yes}	  	
&	\multicolumn{1}{|c|}{yes}		
&	\multicolumn{1}{|c|}{yes}		\\ \cline{3-5}
& {\sl Classical} \T\B 	
&	\multicolumn{1}{|c|}{yes}		
&	\multicolumn{1}{|c|}{no}		
&	\multicolumn{1}{|c|}{yes}		\\ \cline{3-5}
& {\sl Quantum}	\T\B	
&	\multicolumn{1}{|c|}{yes}		
&	\multicolumn{1}{|c|}{yes}		
&	\multicolumn{1}{|c|}{no}		\\ \cline{3-5}
\end{tabular}
\caption{The table describes the conditions of the assignment as a function of system-environment correlations. If we demand an assignment to be linear, consistent, and positive, then the system-environment state must be uncorrelated.  If we give up consistency only,  then the system-environment correlations can be of the classical form.  Finally, if we give up positivity only, we can get quantum correlations for the system-environment state. Note our definitions of quantum and classical correlations do not coincide with the definitions entangled and separable states.}\label{tablesuma}
\end{table}
In response to Pechukas, Alicki proposed three ``natural'' conditions that can be placed on the assignment \cite{PhysRevLett.75.3020}. The assignment map should be

\begin{tabular}{ll}
$i.$ & 
{linear: $\mathcal{A}[a\,\eta_1 + b\,\eta_2] =a\, \mathcal{A}[\eta_1]
+b\, \mathcal{A}[\eta_2]$, }\cr
$ii.$ & 
{consistent: $\mbox{Tr}_\mathsf{E} \left( \mathcal{A}[\eta] \right) =\eta$, }\cr
$iii.$ & {positive: $\mathcal{A}[\eta]\geq0$ for all $\eta$. }\cr
\end{tabular}

\noindent Alicki argued that since the preparation of the system will affect the state of the environment, a linear assignment cannot be well-defined for initially correlated $\mathsf{SE}$ states.  Replying to Alicki, Pechukas agrees to give up linearity and accept \emph{``nonlinearity as a feature of reduced dynamics outside the weak coupling regime"} \cite{PhysRevLett.75.3021}. But, giving up linearity is not desirable; it would disrupt quantum theory in a way that is not experimentally supported \cite{Weinberg89}. A very simple argument favoring the linearity of quantum mechanics was given by Jordan \cite{jordanlinear}. We reexamine the arguments by Pechukas and Alicki against the linearity of assignment maps. In this paper, we study the significance of preserving the linearity condition at the expense of giving up consistency or positivity, as summarized in Table \ref{tablesuma}. 

The properties of the initial system-environment correlations can determine some of the properties of the dynamical map. For example, in \cite{Rodriguez07a} it was shown how some information about total bipartite correlations at the initial time (distinguishing between classical and quantum correlations) could be inferred only from the positivity of the open system dynamics (the sign of its eigenvalues) of one of the parts. This suggest that although the full dynamics do depend on the system-environment correlation together with a particular unitary coupling, the initial correlations (and thus the corresponding assignment map) do fix at least some of the parameters of the dynamics independent of a specific unitary coupling. Understanding the mathematical structure of assignment maps is crucial to the interpretation of open quantum systems as reduced dynamics, and the realm of validity of quantum process tomography as an experimental technique \cite{Kuah07a,Modi09a}.

Our goal in this paper is to show how different types of $\mathsf{SE}$ correlations lead to different conditions for the assignment map. We start by reviewing Pechukas' theorem.  Based on that, we define a \emph{linear assignment}.  We conclude that linearity can be preserved, but the limitations on the validity of the assignment maps come from relaxing positivity or consistency. Finally, we give an operational interpretation to the limitations of a linear assignment in the framework of the quantum no-broadcasting theorem.

\section{Pechukas' Theorem} \label{pechukasproof}
We start by reviewing Pechukas' proof \cite{PhysRevLett.73.1060}.  Pechukas first chose four pure states along different directions: 
\begin{align}
\eta_{1}=\frac{1}{2}\left(\openone+\sigma_x\right),\quad
\eta_{2}=\frac{1}{2}\left(\openone+\sigma_y\right),\nonumber\\
\eta_{4}=\frac{1}{2}\left(\openone-\sigma_x\right),\quad
\eta_{5}=\frac{1}{2}\left(\openone-\sigma_y\right).
\end{align}
He required the assignment to be positive; considering that the states above are \emph{pure}, they cannot be correlated to the states of $\mathsf{E}$ and the action of the assignment must produce a product state in $\mathsf{SE}$. For each $\mathsf{S}$ state $\eta_{i}$ a corresponding $\mathsf{E}$ state $\tau_i$ is assigned by the assignment map such that the total $\mathsf{SE}$ state can be constructed, 
\begin{gather}\label{pechuass}
\eta_{i} \rightarrow \mathcal{A}\left[ \eta_{i} \right]=\eta_{i}\otimes\tau_i.
\end{gather}

Pechukas notes that he can combine the pure states above to produce the maximally mixed state in two different ways, $\frac{1}{2}\eta_{1}+\frac{1}{2}\eta_{4}=\frac{1}{2}\eta_{2}+\frac{1}{2}\eta_{5}=\frac{1}{2}\openone$.
He then applies the assignment to $\frac{1}{2}\openone$, and \emph{by linearity} he obtains, 
\begin{gather}\label{assigned}
\frac{1}{2}\eta_{1}\otimes\tau_1
+\frac{1}{2}\eta_{4}\otimes\tau_4
=\frac{1}{2}\eta_{2}\otimes\tau_2
+\frac{1}{2}\eta_{5}\otimes\tau_5.
\end{gather}
Taking the expectation value of both sides with respect of $\eta_{1}$ yields $2\tau_1=\tau_2+\tau_5$.
Taking similar expectation values with respect to $\eta_{2,3,5}$ lead to a system of equations that made him conclude that $\tau_1=\tau_2=\tau_4=\tau_5$. After this, he similarly defines pure states along the $z-$direction, $\eta_3=\frac{1}{2}\left(\openone+\sigma_z\right) \;\;\;\mbox{and}\;\;\;
\eta_6=\frac{1}{2}\left(\openone-\sigma_z\right)$, and carries out the same argument replacing the states along the $y-$direction with these states along $z$. Pechukas reasoned
by linearity that all states of $\mathsf{S}$ must be assigned to a single $T$, yielding an uncorrelated state in $\mathsf{SE}$.

To define a linear assignment, we note that a generic state in the space of a qubit needs at most four linearly independent matrices to fully describe it. For example, a good choice would be to chose the following projectors to span the qubit space:
\begin{gather}\label{linset}
\mathbf{P}_1=\eta_{1}, \;\;
\mathbf{P}_2=\eta_{2}, \;\;
\mathbf{P}_3=\eta_{3}, \;\;
\mathbf{P}_4=\eta_{4}.
\end{gather}
Any density matrix of a qubit can be written in terms of these projectors as $\eta=\sum_i q_i \mathbf{P}_i$. The decomposition in terms of a set of linearly independent projectors is not a convex decomposition. For example, $\eta_{5}$ can be written as a linear combination of the fixed states in Eq.~(\ref{linset}) as, 
\begin{gather}\label{theotherone}
\eta_{5}=\mathbf{P}_1+\mathbf{P}_4-\mathbf{P}_2
		=\eta_{1}+\eta_{4}-\eta_{2}.
\end{gather}
In other words, $\eta_{5}$ is linearly dependent of the other projectors in Eq.~(\ref{linset}).

Pechukas assumed positivity to show that
\begin{gather}
\eta_{5} \rightarrow \mathcal{A}\left[ \eta_{5} \right]=\eta_{5}\otimes\tau_5,
\end{gather}
with $\tau_5$ independent of the others to prove his theorem. We take a different approach by assuming linearity but not positivity. A map is linear when its action is defined on a set of linearly independent matrices, and preserves linear mixtures of its domain. Applying the assignment linearly to $\eta_5$
should give 
\begin{gather}
\eta_{5}\rightarrow\mathcal{A}[\eta_5]=\eta_{1}\otimes \tau_1 +\eta_{4}\otimes\tau_4 -\eta_{2}\otimes\tau_2, 
\end{gather}
which satisfies Eq.~(\ref{assigned}), but the resulting state is not of product form.

In the next section, we prove Pechukas' theorem by explicitly constructing a linear assignment map, and going beyond the single qubit case.  

\section{Linear assignments}

We define the \emph{most general linear assignment} by its independent action on a fixed (but arbitrary) set of linearly independent set of projectors, $\{\mathbf{P}_i\}$, that span the space of the $\mathsf{S}$,
\begin{gather}\label{linassign}
\mathbf{P}_i\rightarrow \mathcal{A}\left[\mathbf{P}_i \right]=\mathbf{P}_i\otimes \tau_i.
\end{gather}
These projectors span the space of an arbitrarily large quantum system and are not limited to the single qubit case. Any state of $\mathsf{S}$ is given as $\eta =\sum_i q_i \mathbf{P}_i$, with real coefficients $q_i$ such that $\sum_i q_i=1$, but $q_i$ are not necessarily positive. Furthermore, $\tau_i$ are required to be of unit-trace and Hermitian to ensure that the assignment is a trace and Hermiticity preserving map, see Appendix \ref{app1} for proofs.

When we speak of the space $\mathsf{S}$ ($\mathsf{E}$), we refer to the space of operators that act on the Hilbert sub-spaces of the system (environment);  more specifically, we are talking about density operators.  Such operators can be spanned by matrices that form a linearly independent basis, \emph{e.g.} the generators of $SU(N)$ group (see \cite{TilmaSudarshan02} and the references within). We construct the linearly independent basis using positive, Hermitian, unit-trace, rank one projectors, \emph{i.e.} linearly independent pure states.  The advantage of choosing this basis is that the each of the element is a positive matrix, a property that we exploit many times along this paper. Since all possible states of $\mathsf{S}$ can be written as a linear sum of projectors, $\{\mathbf{P}_i\}$, the assignment satisfies the linearity condition, i.e. $\mathcal{A}[\sum q_i \mathbf{P}_i]=\sum q_i\mathcal{A}[\mathbf{P}_i]$.  It also satisfies the consistency condition due to $\mbox{Tr}[\tau_i]=1$, $\mbox{Tr}_{\mathsf{E}}[A[\eta]]=\eta$.  We examine the positivity condition next.

{\sl Lemma 1.} 
If the linear assignment is positive, then each matrix $\tau_i$ must be positive.

{\sl Proof.} 
Suppose the assignment is positive, $\mathcal{A}[\mathbf{P}_i]\geq 0$.  Which means $\mathbf{P}_i\otimes\tau_i\geq0.$ And since $\mathbf{P}_i\geq0$, for $\mathcal{A}$ to be positive $\tau_i\geq0$. $\square$

{\sl Remark.} 
The converse of the last lemma is not true.  That is, if all $\tau_i\geq0$ does not mean that $\mathcal{A}[\eta]\geq0.$  We will show this later by an explicit example.

{\sl Theorem 1.}  
A linear assignment, satisfying conditions $(i)$ and $(ii)$, will also satisfy condition $(iii)$ if and only if it assigns a single state, $T$, to  all projectors that span the space of $\mathsf{S}$.

{\sl Proof.} 
We want to show that if the assignment is positive for all $\eta$, then all $\tau_i$ are the same, $\{\tau_i\} =T$. We begin with ``only if" direction.  Assume $\tau_i=T$  for all  $i$, then $\mathcal{A}[\mathbf{P}_i]=\mathbf{P}_i\otimes T$ for all $i$. The action of the assignment on a generic state, $\eta$, is, 
\begin{gather}
\mathcal{A}[\eta]=\mathcal{A}[\sum_i q_i \mathbf{P}_i]=\sum_i q_i \mathbf{P}_i\otimes {T}=\eta\otimes{T}\geq0,
\end{gather}
for all $\eta\geq0$ and ${T}\geq0$.  The set $\eta\geq0$ is the set of all states hence the assignment is positive.
	
Now to prove the ``if" direction, assume that the linear assignment is positive for all states in its domain. Consider the action of the assignment on an arbitrary pure state $\mathbf{R}$. Since $\mathbf{R}$ is a pure state, the result of the action of the assignment has to be a state in the product form, $\mathcal{A}[\mathbf{R}]=\mathbf{R}\otimes {T}.$ Note that any state of the system can be represented in terms of the fixed set of projectors \footnote{Since $\mathbf{R}$ is pure, there are stringent conditions on the values $\{q_i\}$. These conditions are not easy to write down in general.  For a qubit, the $\{q_i\}$ in terms of the Bloch vector parameters are $q_1=\frac{1}{2}(1+a_1-a_2-a_3)$, $q_2=a_2$, $q_3=a_3$, and $q_4=\frac{1}{2}(1-a_1-a_2-a_3)$ for the choice of linearly independent projectors in Eq.~(\ref{linset}).  Additionally, they satisfy $a_1^2+a_2^2+a_3^2=1$ for pure states.}: $\mathbf{R} =\sum_i q_i \mathbf{P}_i$.  Substituting this for $\mathbf{R}$ before and after the action of assignment gives us two sets of linearly independent equations.  Applying the assignment and then substituting for $\mathbf{R}=\sum_iq_i\mathbf{P}_i$ gives,
\begin{align}\label{eq14}
\mathcal{A}[\mathbf{R}]
&=\sum_i q_i \mathbf{P}_i\otimes {T}
\Longleftrightarrow \mathcal{A}[\mathbf{P}_i]=\mathbf{P}_i\otimes{T}.
\end{align}
While applying linear assignment after substituting $\mathbf{R}=\sum_iq_i\mathbf{P}_i$ gives,
\begin{align}\label{eq15}
\mathcal{A}[\mathbf{R}]
&=\sum_iq_i\mathbf{P}_i\otimes\tau_i,
\Longleftrightarrow
\mathcal{A}\left[\mathbf{P}_i \right]
=\mathbf{P}_i\otimes \tau_i.
\end{align}
Matching the linearly independent terms of Eqs.~(\ref{eq14}) and (\ref{eq15}) we obtain $\mathbf{P}_i\otimes T =\mathbf{P}_i\otimes\tau_i$ for all $i$. Taking the trace with respect to the system gives that $\mathbf{T}=\tau_i$ for all $i$. $\square$

{\sl Remark.} 
The theorem above says that if one demands that a linear assignment also be positive and consistent, then the only valid assignment is one that yields no correlations between $\mathsf{SE}$.  The result is simply a tensor product of the states of $\mathsf{S}$ and a single state of $\mathsf{E}$, which agrees with Pechukas' theorem. 

Theorem 1 suggests that if we enforce all conditions for the assignment simultaneously then the assignment leads to an uncorrelated states of $\mathsf{SE}$. Are we then forced to agree the conclusion of Pechukas and Alicki and accept nonlinearity \cite{PhysRevLett.75.3021}? In the next two subsections, we argue that this is not the case and discuss how relaxing the positivity or the consistency conditions of the assignment map is more reasonable. We start by relaxing only the consistency condition. Then, we will discuss relaxing only the positivity condition.

\subsection{Relaxing consistency}

Let us consider the situation where all initial states of $\mathsf{S}$ are projected into orthogonal states. In this case, each orthogonal state interacts with the environment through a separate quantum channel.  The linear assignment relevant to this physical situation has the form:
\begin{gather}\label{zdassign}
\mathcal{A}\left[\eta\right]= \sum_i \mbox{Tr}\left[\eta\Pi_i\right] \; \Pi_i\otimes\tau_i,
\end{gather}
where $\{\Pi_i\}$ are a set of orthonormal projectors on the space of $\mathsf{S}$ \footnote{The orthonormal projectors should not to be confused with the set of linearly independent projectors defined earlier.}. The state on the r.h.s of Eq.~(\ref{zdassign}) is \emph{classically correlated}, meaning it has zero quantum discord \cite{henderson01a,PhysRevLett.88.017901}, and has a deep connection to completely positive maps as studied by us in \cite{Rodriguez07a}, and extended by Shabani and Lidar \cite{shabanilidar09a}.

The assignment in Eq.~(\ref{zdassign}) is a subclass of the assignment from Eq.~(\ref{linassign}) and thus it is linear, Hermitian, and trace preserving.  We now prove that this assignment violates the consistency condition but is still positive. This case was initially suggested by Alicki \cite{PhysRevLett.75.3020}.

{\sl Theorem 2.} 
The assignment in Eq.~(\ref{zdassign}) is not consistent.

{\sl Proof.} 
We can prove this by direct computation: 
\begin{align}
\mbox{Tr}_\mathsf{E}\left[\mathcal{A}[\eta]\right]
=&\sum_i\mbox{Tr}\left[\eta\Pi_i\right]\Pi_i\mbox{Tr}_\mathsf{E} \left[\tau_i\right]\nonumber\\
=&\sum_i\mbox{Tr}\left[\eta\Pi_i\right]\Pi_i\neq\eta.
\end{align}
The assignment is only consistent when $\eta=\sum p_i \Pi_i$, where $p_i\ge 0$ and $\sum_i p_i=1$. In other words, it is consistent for states diagonal in the basis defined by the projectors, $\{\Pi_i\}$.  $\square$

{\sl Theorem 3.} 
The assignment in Eq.~(\ref{zdassign}) is positive if and only if $\tau_i\geq0.$

{\sl Proof.} 
Note that, by definition, $\Pi_i\geq0$ and also $\mbox{Tr}[\eta\Pi_i]\geq0$ for all $\eta\geq0$.  Then the terms $\Pi_i\otimes\tau_i\geq0$ for all $\tau_i\geq0$.  The convex combination of positive terms if positive, and hence the assignment is positive.

On the other hand, the relationship $\Pi_i\otimes\tau_i<0$ is true for all $\tau_i<0$. Each $\Pi_i\otimes\tau_i$ are orthogonal to each other; they are block diagonal and thus the eigenvalues of each block are the eigenvalues of $\tau_i$ multiplied by $\mbox{Tr}[\eta\Pi_i]$, (see \cite{modigeo} for proof).  Therefore, if $\tau_i<0$ then the total state has a negative eigenvalue and the assignment is not positive. $\square$

{\sl Remark.} 
This assignment is positive and contains some correlations between the system and environment. However, it is unable to reach all possible $\mathsf{SE}$ states as it only outputs classically correlated states. The downside, of course, is that the trace with respect to $\mathsf{E}$ gives back $\eta$ only when the state is in the eigenbasis $\{\Pi_i\}$.  The conclusion is that a positive linear assignment is either allowed only for uncorrelated states or does not abide to the consistency condition.

\subsection{Relaxing positivity}\label{compdomain}
The positivity requirement is fundamentally inconsistent with having correlations of $\mathsf{SE}$. By definition, correlations imply that not all states of a subpart may be compatible \cite{jordan:052110}. The correlations defined by the map constrain the domain of $\mathsf{S}$, meaning that certain $\{q_i\}$, which lead to a valid state of $\mathsf{S}$, may not lead to a valid state of $\mathsf{SE}$.

More generally, we can say that if an assignment is linear and consistent, but not positive, then there must be a compatibility domain.  The compatibility domain was defined by Jordan, Shaji, and Sudarshan \cite{jordan:052110}, and here we interpret it as the subset of density matrices of $\mathsf{S}$ that are mapped by the assignment to valid density matrices in $\mathsf{SE}$.  The linear assignment in Eq.~(\ref{linassign}) is a trace and Hermiticity preserving map, therefore the compatibility domain is a function only of the positivity condition of the map.  Let us illustrate this with a simple example.

Consider a $\mathsf{SE}$, 
$\rho=\sum_i q_i \mathbf{P}_i\otimes\mathbf{\Xi}_i$,
where $\{\mathbf{\Xi}_i\}$ form a complete set of orthonormal projectors in the space of $\mathsf{E}$. Now suppose this is the initial state of $\mathsf{SE}$ onto which we want to define a linear assignment.  We can do that by defining the assignment as
$\mathcal{A}[\mathbf{P}_i]=\mathbf{P}_i\otimes\mathbf{\Xi}_i$.
We can immediately write down the action of the assignment on a generic state of the system, 
\begin{gather}
\mathcal{A}\left[\eta\right]=\mathcal{A}\left[\sum_i q_i\mathbf{P}_i\right]=\sum_i q_i\mathbf{P}_i\otimes\mathbf{\Xi}_i.
\end{gather}
Note that not all choices of $q_i$ will yield a positive state.  Since $\{\mathbf{\Xi}_i\}$ form a complete orthonormal set, all $q_i$ are the eigenvalues of the state of $\mathsf{E}$,  which means that only the system states with $q_i\geq0$ are valid set of states for this assignment.  But, $q_i\geq0$ are not the set of all system states, and hence the assignment above is not a positive assignment, and its compatibility domain is the set of states that can be written as $\eta=\sum q_i \mathbf{P}_i$ for all $q_i\geq 0$. This example also illustrates that the converse of \emph{Lemma 1} is not true. That is, if all $\{\tau_i \}$ are positive (for our example $\Xi_i\geq0$) that does not mean that the assignment is positive.

\section{No-cloning and no-broadcasting theorems}

Though the notion of an assignment is completely mathematical, it has deep physical consequences, imposing limitations on the experimentally accessible dynamics and, ultimately, the mathematical structure of quantum mechanics. These consequences are clear when the assignment is analyzed in light of the 
no-cloning theorem \cite{wootterszurek,dieks} and the no-broadcasting theorem \cite{Barnum96a,amir}.  

Simply put, the no-cloning theorem says that the linearity of quantum mechanics implies that \emph{pure} quantum states of the form $|\psi\rangle\langle \psi \vert$ cannot be copied. This can be stated as an argument favoring linear assignments. A cloning map would have the property $\mathcal{C}[|\psi\rangle\langle \psi \vert]
=|\psi\rangle\langle \psi \vert\otimes|\psi\rangle\langle \psi \vert$, which is clearly not linear. Since the no-cloning theorem says that the only states that can be cloned are pure orthogonal states, consistency and linearity would both have to be relaxed.

On the other hand, the no-broadcasting theorem says there is no general linear completely positive map that acting on a general state $\eta=\sum_i q_i\mathbf{P}_i$ (not necessarily pure) can give a bipartite state in $\mathsf{SE}$ such that each of its reduced subparts $\mathsf{S}$ and $\mathsf{E}$ are also $\eta$. More specifically, it says that the only states that can be broadcast are commuting states. We can study the conditions of such a broadcasting map by defining it as a class of assignment maps: a broadcasting assignment map $\mathcal{B}$ is defined as $\mathcal{B}\left[ \eta \right]=\rho$ such that it follows the broadcasting condition $ \mbox{Tr}_\mathsf{E}[\rho] = \mbox{Tr}_\mathsf{S}[\rho]=\eta$. 

The conditions for the broadcasting map are related to the conditions of the assignment.  The broadcasting condition automatically implies the consistency condition. Also, unlike the cloning map, \emph{a broadcasting map can be linear}. We show this by construction: $\mathcal{B}[\mathbf{P}_i]
=\mathbf{P}_i \otimes \mathbf{P}_i$, which is a special class of maps from Eq.~(\ref{linassign}). The action of such a broadcasting map can be defined as,
\begin{equation}\label{broadcast}
\mathcal{B}[\eta]
=\mathcal{B}\left[\sum_i q_i\mathbf{P}_i\right]
=\sum_i q_i\mathbf{P}_i \otimes \mathbf{P}_i,
\end{equation} which, by taking the trace on each side, fulfills the broadcasting condition for a linear map. It is the positivity condition that cannot be imposed: if the broadcasting map acts on a general state $\eta$ it might yield a matrix $\rho$ that does not have positive eigenvalues and is not a density matrix with a physical interpretation. In general the broadcasted state has correlations in $\mathsf{SE}$. Therefore, it is unreasonable to assume that all valid $\mathsf{S}$ states will be compatible with $\mathsf{SE}$ states.

Consider the following example that follows closely our argument from Section~\ref{pechukasproof} and uses the density matrices defined in Eq.~(\ref{linset}) and Eq.~(\ref{theotherone}). First, note that $\eta_1$ and $\eta_4$ are commuting matrices, and can be broadcasted by Eq.~(\ref{broadcast}) such that: $\mathcal{B}[\eta_1]
=\eta_1\otimes \eta_1$ and $\mathcal{B}[\eta_4]
=\eta_4\otimes \eta_4$, which certainly are valid density matrices in $\mathsf{SE}$. The map could also broadcast a state that does not commute with $\eta_1$ and $\eta_4$, such as $\eta_2$, $\mathcal{B}[\eta_2]
=\eta_2\otimes \eta_2$, which is also a valid density matrix in $\mathsf{SE}$. It is when we try to broadcast the state $\eta_5$ that positivity is violated. By linearity of Eq.~(\ref{broadcast}) and the decomposition from Eq.~(\ref{theotherone}) we obtain the matrix 
\begin{gather}
\mathcal{B}[\eta_5]
=\mathbf{P}_1 \otimes \mathbf{P}_1+\mathbf{P}_4 \otimes \mathbf{P}_4 -\mathbf{P}_2 \otimes \mathbf{P}_2,
\end{gather}
which does follow the broadcasting condition, but has negative eigenvalues. Thus, the no-broadcasting condition comes from the negativity of the broadcasting assignment maps. Since this map can be positive on a subset of states, it has a compatibility domain as discussed in Section \ref{compdomain}.

Similarly, we can think of the assignment in Eq.~(\ref{zdassign}) also as broadcasting map, but only for classical information, 
$\mathcal{A}[\eta]=\sum_i \mbox{Tr}[\eta\Pi_i]\Pi_i\otimes\Pi_i$.
This map sends any information of $\eta$ that is diagonal in the basis given by $\{\Pi_i\}$ from $\mathsf{S}$ to $\mathsf{E}$.  This is in accordance with the no-broadcasting theorem, since commuting states can be broadcasted.  The map above goes further and shows that such operations can broadcast partial information from states that do not commute, i.e. states that are not diagonal in basis $\{\Pi_i\}$.

The analysis in this section suggests that linear assignments can be interpreted as generalized broadcasting from $\mathsf{S}$ to $\mathsf{E}$, regardless of the size of $\mathsf{E}$.  This gives an operational meaning to the mathematical concept of assignments on a physical basis.

\section{Conclusion}
We have considered the consequences of relaxing consistency or positivity of linear assignment maps. First, we show how an assignment map cannot be linear, positive, consistent, and have correlations. We show that, by giving up consistency, the assignment map can have classical correlations, and be linear and positive. Giving up positivity allows quantum correlations for a linear and consistent assignment map. The physical intuition of assignment maps is shown to be related to the no-broadcasting theorem. The no-broadcasting condition comes from the positivity condition of the assignment map, not from its linearity.

\acknowledgements

We thank A. Kuah for introducing us to this problem.  We are grateful to E.C.G. Sudarshan for helpful discussions.  K.M. thanks Wonmin Son for helpful comments,  acknowledges the financial supported by the National Research Foundation and the Ministry of Education of Singapore, and the hospitality of Department of Chemistry and Chemical Biology at Harvard University. C.A.R. thanks the Mary-Fieser Postdoctoral Fellowship program and the hospitality of the Centre for Quantum Technologies at the National University of Singapore. This material is based upon work supported as part of the Center for Excitonics, an Energy Frontier Research Center funded by the U.S. Department of Energy, Office of Science, Office of Basic Energy Sciences under Award Number DE-SC0001088.

\appendix
\section{Hermiticity and trace preservation of assignments}\label{app1}

{\sl Proposition 1.} 
The linear assignment in Eq.~(\ref{linassign}) is a Hermiticity preserving map if and only if the matrices of $\mathsf{E}$, $\tau_i$, are Hermitian.

{\sl Proof.} 
If $\tau_i=\tau_i^\dag$ and using the fact $\mathbf{P}_i=\mathbf{P}_i^\dag$ we get
$\left(\mathcal{A}[\mathbf{P}_i]\right)^\dag =\left(\mathbf{P}_i \otimes \tau_i\right)^\dag = \mathcal{A}[\mathbf{P}_i].$ To prove the other direction, assume $\mathcal{A}[\mathbf{P}_i] =\left(\mathcal{A}[\mathbf{P}_i]\right)^\dag$,  which leads to $\mathbf{P}_i\otimes\tau_i = \mathbf{P}_i\otimes\tau_i^\dag$. Taking the trace with respect to $\mathsf{S}$ yields $\tau_i=\tau_i^\dag.$ $\square$

{\sl Proposition 2.} 
The linear assignment in Eq.~(\ref{linassign}) is a trace preserving map, i.e. $\mbox{Tr}[\mathcal{A}[\eta]]=\mbox{Tr}[\eta]$, if and only if $\tau_i$ are unit-trace.

{\sl Proof.} 
The set of all states contain the linearly independent projectors, so let us only look at the action of the map on those.
\begin{gather}
\mbox{Tr}\left[\mathcal{A}[\mathbf{P}_i]\right]
=\mbox{Tr}[\mathbf{P}_i\otimes\tau_1]
=\mbox{Tr}[\mathbf{P}_i] \times \mbox{Tr}[\tau_i]
=\mbox{Tr}[\mathbf{P}_i].\nonumber
\end{gather}
Since $\mbox{Tr}[\mathbf{P}_i]=1$, the assignment is trace preserving if and only if $\mbox{Tr}[\tau_i]=1$, and by linearity we have $\mbox{Tr}[\mathcal{A}[\eta]]=1.$ $\square$

\bibliography{pechukas.bib}

\begin{thebibliography}{38}
\expandafter\ifx\csname natexlab\endcsname\relax\def\natexlab#1{#1}\fi
\expandafter\ifx\csname bibnamefont\endcsname\relax
  \def\bibnamefont#1{#1}\fi
\expandafter\ifx\csname bibfnamefont\endcsname\relax
  \def\bibfnamefont#1{#1}\fi
\expandafter\ifx\csname citenamefont\endcsname\relax
  \def\citenamefont#1{#1}\fi
\expandafter\ifx\csname url\endcsname\relax
  \def\url#1{\texttt{#1}}\fi
\expandafter\ifx\csname urlprefix\endcsname\relax\def\urlprefix{URL }\fi
\providecommand{\bibinfo}[2]{#2}
\providecommand{\eprint}[2][]{\url{#2}}

\bibitem[{\citenamefont{Sudarshan et~al.}(1961)\citenamefont{Sudarshan,
  Mathews, and Rau}}]{PhysRev.121.920}
\bibinfo{author}{\bibfnamefont{E.~C.~G.} \bibnamefont{Sudarshan}},
  \bibinfo{author}{\bibfnamefont{P.~M.} \bibnamefont{Mathews}},
  \bibnamefont{and} \bibinfo{author}{\bibfnamefont{J.}~\bibnamefont{Rau}},
  \bibinfo{journal}{Phys. Rev.} \textbf{\bibinfo{volume}{121}},
  \bibinfo{pages}{920} (\bibinfo{year}{1961}).

\bibitem[{\citenamefont{Jordan and Sudarshan}(1961)}]{jordan:772}
\bibinfo{author}{\bibfnamefont{T.~F.} \bibnamefont{Jordan}} \bibnamefont{and}
  \bibinfo{author}{\bibfnamefont{E.~C.~G.} \bibnamefont{Sudarshan}},
  \bibinfo{journal}{Journal of Mathematical Physics}
  \textbf{\bibinfo{volume}{2}}, \bibinfo{pages}{772} (\bibinfo{year}{1961}).

\bibitem[{\citenamefont{Choi}(1972)}]{choi72a}
\bibinfo{author}{\bibfnamefont{M.~D.} \bibnamefont{Choi}},
  \bibinfo{journal}{Can. J. Math.} \textbf{\bibinfo{volume}{24}},
  \bibinfo{pages}{520} (\bibinfo{year}{1972}).

\bibitem[{\citenamefont{Choi}(1975)}]{choi75}
\bibinfo{author}{\bibfnamefont{M.~D.} \bibnamefont{Choi}},
  \bibinfo{journal}{Linear Algebra and Appl.} \textbf{\bibinfo{volume}{10}},
  \bibinfo{pages}{285} (\bibinfo{year}{1975}).

\bibitem[{\citenamefont{Simmons and Park}(1980)}]{Simmons80a}
\bibinfo{author}{\bibfnamefont{R.}~\bibnamefont{Simmons}} \bibnamefont{and}
  \bibinfo{author}{\bibfnamefont{J.}~\bibnamefont{Park}},
  \bibinfo{journal}{Foundations of Physics} \textbf{\bibinfo{volume}{11}},
  \bibinfo{pages}{47} (\bibinfo{year}{1980}).

\bibitem[{\citenamefont{Raggio and Primas}(1981)}]{raggio}
\bibinfo{author}{\bibfnamefont{G.~A.} \bibnamefont{Raggio}} \bibnamefont{and}
  \bibinfo{author}{\bibfnamefont{H.}~\bibnamefont{Primas}},
  \bibinfo{journal}{Foundations of Physics} \textbf{\bibinfo{volume}{12}},
  \bibinfo{pages}{433} (\bibinfo{year}{1981}).

\bibitem[{\citenamefont{Simmons and Park}(1981)}]{simmonsparkreply}
\bibinfo{author}{\bibfnamefont{R.}~\bibnamefont{Simmons}} \bibnamefont{and}
  \bibinfo{author}{\bibfnamefont{J.}~\bibnamefont{Park}},
  \bibinfo{journal}{Foundations of Physics} \textbf{\bibinfo{volume}{12}},
  \bibinfo{pages}{437} (\bibinfo{year}{1981}).

\bibitem[{\citenamefont{Pechukas}(1994)}]{PhysRevLett.73.1060}
\bibinfo{author}{\bibfnamefont{P.}~\bibnamefont{Pechukas}},
  \bibinfo{journal}{Phys. Rev. Lett.} \textbf{\bibinfo{volume}{73}},
  \bibinfo{pages}{1060} (\bibinfo{year}{1994}).

\bibitem[{\citenamefont{Alicki}(1995)}]{PhysRevLett.75.3020}
\bibinfo{author}{\bibfnamefont{R.}~\bibnamefont{Alicki}},
  \bibinfo{journal}{Phys. Rev. Lett.} \textbf{\bibinfo{volume}{75}},
  \bibinfo{pages}{3020} (\bibinfo{year}{1995}).

\bibitem[{\citenamefont{Pechukas}(1995)}]{PhysRevLett.75.3021}
\bibinfo{author}{\bibfnamefont{P.}~\bibnamefont{Pechukas}},
  \bibinfo{journal}{Phys. Rev. Lett.} \textbf{\bibinfo{volume}{75}},
  \bibinfo{pages}{3021} (\bibinfo{year}{1995}).

\bibitem[{\citenamefont{\ifmmode \check{S}\else
  \v{S}\fi{}telmachovi\ifmmode~\check{c}\else \v{c}\fi{} and
  Bu\ifmmode~\check{z}\else \v{z}\fi{}ek}(2001)}]{PhysRevA.64.062106}
\bibinfo{author}{\bibfnamefont{P.}~\bibnamefont{\ifmmode \check{S}\else
  \v{S}\fi{}telmachovi\ifmmode~\check{c}\else \v{c}\fi{}}} \bibnamefont{and}
  \bibinfo{author}{\bibfnamefont{V.}~\bibnamefont{Bu\ifmmode~\check{z}\else
  \v{z}\fi{}ek}}, \bibinfo{journal}{Phys. Rev. A}
  \textbf{\bibinfo{volume}{64}}, \bibinfo{pages}{062106}
  (\bibinfo{year}{2001}).

\bibitem[{\citenamefont{Jordan et~al.}(2004)\citenamefont{Jordan, Shaji, and
  Sudarshan}}]{jordan:052110}
\bibinfo{author}{\bibfnamefont{T.~F.} \bibnamefont{Jordan}},
  \bibinfo{author}{\bibfnamefont{A.}~\bibnamefont{Shaji}}, \bibnamefont{and}
  \bibinfo{author}{\bibfnamefont{E.~C.~G.} \bibnamefont{Sudarshan}},
  \bibinfo{journal}{Phys. Rev. A} \textbf{\bibinfo{volume}{70}},
  \bibinfo{eid}{052110} (\bibinfo{year}{2004}).

\bibitem[{\citenamefont{Shaji and Sudarshan}(2005)}]{Shaji05a}
\bibinfo{author}{\bibfnamefont{A.}~\bibnamefont{Shaji}} \bibnamefont{and}
  \bibinfo{author}{\bibfnamefont{E.~C.~G.} \bibnamefont{Sudarshan}},
  \bibinfo{journal}{Phys. Lett. A} \textbf{\bibinfo{volume}{341}},
  \bibinfo{pages}{48} (\bibinfo{year}{2005}).

\bibitem[{\citenamefont{Jordan et~al.}(2006)\citenamefont{Jordan, Shaji, and
  Sudarshan}}]{jordan:012106}
\bibinfo{author}{\bibfnamefont{T.~F.} \bibnamefont{Jordan}},
  \bibinfo{author}{\bibfnamefont{A.}~\bibnamefont{Shaji}}, \bibnamefont{and}
  \bibinfo{author}{\bibfnamefont{E.~C.~G.} \bibnamefont{Sudarshan}},
  \bibinfo{journal}{Phys. Rev. A} \textbf{\bibinfo{volume}{73}},
  \bibinfo{eid}{012106} (pages~\bibinfo{numpages}{9}) (\bibinfo{year}{2006}).

\bibitem[{\citenamefont{Ziman}(2006)}]{ziman06}
\bibinfo{author}{\bibfnamefont{M.}~\bibnamefont{Ziman}},
  \bibinfo{journal}{quant-ph/0603166}  (\bibinfo{year}{2006}).

\bibitem[{\citenamefont{Rodriguez-Rosario
  et~al.}(2008)\citenamefont{Rodriguez-Rosario, Modi, Kuah, Shaji, and
  Sudarshan}}]{Rodriguez07a}
\bibinfo{author}{\bibfnamefont{C.~A.} \bibnamefont{Rodriguez-Rosario}},
  \bibinfo{author}{\bibfnamefont{K.}~\bibnamefont{Modi}},
  \bibinfo{author}{\bibfnamefont{A.}~\bibnamefont{Kuah}},
  \bibinfo{author}{\bibfnamefont{A.}~\bibnamefont{Shaji}}, \bibnamefont{and}
  \bibinfo{author}{\bibfnamefont{E.~C.~G.} \bibnamefont{Sudarshan}},
  \bibinfo{journal}{J. Phys. A: Math. Theor} \textbf{\bibinfo{volume}{41}},
  \bibinfo{pages}{205301} (\bibinfo{year}{2008}).

\bibitem[{\citenamefont{Kuah et~al.}(2007)\citenamefont{Kuah, Modi, Rodriguez,
  and Sudarshan}}]{Kuah07a}
\bibinfo{author}{\bibfnamefont{A.}~\bibnamefont{Kuah}},
  \bibinfo{author}{\bibfnamefont{K.}~\bibnamefont{Modi}},
  \bibinfo{author}{\bibfnamefont{C.~A.} \bibnamefont{Rodriguez}},
  \bibnamefont{and} \bibinfo{author}{\bibfnamefont{E.~C.~G.}
  \bibnamefont{Sudarshan}}, \bibinfo{journal}{Phys. Rev. A}
  \textbf{\bibinfo{volume}{76}}, \bibinfo{eid}{0706.0394}
  (\bibinfo{year}{2007}).

\bibitem[{\citenamefont{Carteret et~al.}(2008)\citenamefont{Carteret, Terno,
  and Zyczkoski}}]{Carteret08a}
\bibinfo{author}{\bibfnamefont{H.}~\bibnamefont{Carteret}},
  \bibinfo{author}{\bibfnamefont{D.}~\bibnamefont{Terno}}, \bibnamefont{and}
  \bibinfo{author}{\bibfnamefont{K.}~\bibnamefont{Zyczkoski}},
  \bibinfo{journal}{Phys. Rev. A} \textbf{\bibinfo{volume}{77}},
  \bibinfo{pages}{042113} (\bibinfo{year}{2008}).

\bibitem[{\citenamefont{Rodriguez-Rosario and Sudarshan}(2008)}]{Rodriguez08}
\bibinfo{author}{\bibfnamefont{C.~A.} \bibnamefont{Rodriguez-Rosario}}
  \bibnamefont{and} \bibinfo{author}{\bibfnamefont{E.~C.~G.}
  \bibnamefont{Sudarshan}}, \bibinfo{journal}{arXiv:0803.1183}
  (\bibinfo{year}{2008}).

\bibitem[{\citenamefont{Shabani and
  Lidar}(2009{\natexlab{a}})}]{shabanilidar09a}
\bibinfo{author}{\bibfnamefont{A.}~\bibnamefont{Shabani}} \bibnamefont{and}
  \bibinfo{author}{\bibfnamefont{D.~A.} \bibnamefont{Lidar}},
  \bibinfo{journal}{Phys. Rev. Lett.} \textbf{\bibinfo{volume}{102}},
  \bibinfo{pages}{100402} (\bibinfo{year}{2009}{\natexlab{a}}).

\bibitem[{\citenamefont{Shabani and
  Lidar}(2009{\natexlab{b}})}]{shabanilidar09b}
\bibinfo{author}{\bibfnamefont{A.}~\bibnamefont{Shabani}} \bibnamefont{and}
  \bibinfo{author}{\bibfnamefont{D.~A.} \bibnamefont{Lidar}},
  \bibinfo{journal}{Phys. Rev. A} \textbf{\bibinfo{volume}{80}},
  \bibinfo{pages}{012309} (\bibinfo{year}{2009}{\natexlab{b}}).

\bibitem[{\citenamefont{Modi and Sudarshan}(2009)}]{Modi09b}
\bibinfo{author}{\bibfnamefont{K.}~\bibnamefont{Modi}} \bibnamefont{and}
  \bibinfo{author}{\bibfnamefont{E.~C.~G.} \bibnamefont{Sudarshan}},
  \bibinfo{journal}{arXiv.org:0904.4663}  (\bibinfo{year}{2009}).

\bibitem[{\citenamefont{Shaji}(2005)}]{shaji05dis}
\bibinfo{author}{\bibfnamefont{A.}~\bibnamefont{Shaji}}, Ph.D. thesis,
  \bibinfo{school}{The University of Texas at Austin} (\bibinfo{year}{2005}).

\bibitem[{\citenamefont{Henderson and Vedral}(2001)}]{henderson01a}
\bibinfo{author}{\bibfnamefont{L.}~\bibnamefont{Henderson}} \bibnamefont{and}
  \bibinfo{author}{\bibfnamefont{V.}~\bibnamefont{Vedral}},
  \bibinfo{journal}{J. Phys. A} \textbf{\bibinfo{volume}{34}},
  \bibinfo{pages}{6899} (\bibinfo{year}{2001}).

\bibitem[{\citenamefont{Ollivier and Zurek}(2001)}]{PhysRevLett.88.017901}
\bibinfo{author}{\bibfnamefont{H.}~\bibnamefont{Ollivier}} \bibnamefont{and}
  \bibinfo{author}{\bibfnamefont{W.~H.} \bibnamefont{Zurek}},
  \bibinfo{journal}{Phys. Rev. Lett.} \textbf{\bibinfo{volume}{88}},
  \bibinfo{pages}{017901} (\bibinfo{year}{2001}).

\bibitem[{\citenamefont{Rodr\'{i}guez-Rosario}(2008)}]{rodriguezdis}
\bibinfo{author}{\bibfnamefont{C.~A.} \bibnamefont{Rodr\'{i}guez-Rosario}},
  Ph.D. thesis, \bibinfo{school}{The University of Texas at Austin}
  (\bibinfo{year}{2008}).

\bibitem[{\citenamefont{Modi}(2009)}]{Modi09a}
\bibinfo{author}{\bibfnamefont{K.}~\bibnamefont{Modi}},
  \bibinfo{journal}{arXiv:0903.2027}  (\bibinfo{year}{2009}).

\bibitem[{\citenamefont{Modi}(2008{\natexlab{a}})}]{Modidis}
\bibinfo{author}{\bibfnamefont{K.}~\bibnamefont{Modi}}, Ph.D. thesis,
  \bibinfo{school}{The University of Texas at Austin}
  (\bibinfo{year}{2008}{\natexlab{a}}).

\bibitem[{\citenamefont{Zurek}(2003)}]{zurekdecoherence}
\bibinfo{author}{\bibfnamefont{W.}~\bibnamefont{Zurek}},
  \bibinfo{journal}{Review of Modern Physics} \textbf{\bibinfo{volume}{75}},
  \bibinfo{pages}{715} (\bibinfo{year}{2003}).

\bibitem[{\citenamefont{Abrams and Lloyd}(1998)}]{Abrams98a}
\bibinfo{author}{\bibfnamefont{D.}~\bibnamefont{Abrams}} \bibnamefont{and}
  \bibinfo{author}{\bibfnamefont{S.}~\bibnamefont{Lloyd}},
  \bibinfo{journal}{Phys. Rev. Lett.} \textbf{\bibinfo{volume}{81}},
  \bibinfo{pages}{3992} (\bibinfo{year}{1998}).

\bibitem[{\citenamefont{Weinberg}(1989)}]{Weinberg89}
\bibinfo{author}{\bibfnamefont{S.}~\bibnamefont{Weinberg}},
  \bibinfo{journal}{Phys. Rev. Lett.} \textbf{\bibinfo{volume}{62}},
  \bibinfo{pages}{485} (\bibinfo{year}{1989}).

\bibitem[{\citenamefont{Jordan}(2007)}]{jordanlinear}
\bibinfo{author}{\bibfnamefont{T.~F.} \bibnamefont{Jordan}},
  \bibinfo{journal}{arXiv:quant-ph/070217v1}  (\bibinfo{year}{2007}).

\bibitem[{\citenamefont{Tilma and Sudarshan}(2002)}]{TilmaSudarshan02}
\bibinfo{author}{\bibfnamefont{T.}~\bibnamefont{Tilma}} \bibnamefont{and}
  \bibinfo{author}{\bibfnamefont{E.~C.~G.} \bibnamefont{Sudarshan}},
  \bibinfo{journal}{J. Phys. A} \textbf{\bibinfo{volume}{35}},
  \bibinfo{pages}{10467} (\bibinfo{year}{2002}).

\bibitem[{\citenamefont{Modi}(2008{\natexlab{b}})}]{modigeo}
\bibinfo{author}{\bibfnamefont{K.}~\bibnamefont{Modi}},
  \bibinfo{journal}{arXiv:0902.0735}  (\bibinfo{year}{2008}{\natexlab{b}}).

\bibitem[{\citenamefont{Wootters and Zurek}(1982)}]{wootterszurek}
\bibinfo{author}{\bibfnamefont{W.}~\bibnamefont{Wootters}} \bibnamefont{and}
  \bibinfo{author}{\bibfnamefont{W.}~\bibnamefont{Zurek}},
  \bibinfo{journal}{Nature} \textbf{\bibinfo{volume}{299}},
  \bibinfo{pages}{802} (\bibinfo{year}{1982}).

\bibitem[{\citenamefont{Dieks}(1982)}]{dieks}
\bibinfo{author}{\bibfnamefont{D.}~\bibnamefont{Dieks}},
  \bibinfo{journal}{Phys. Lett. A} \textbf{\bibinfo{volume}{92}},
  \bibinfo{pages}{271} (\bibinfo{year}{1982}).

\bibitem[{\citenamefont{Barnum et~al.}(1996)\citenamefont{Barnum, Caves, Fuchs,
  Jozsa, and Schumacher}}]{Barnum96a}
\bibinfo{author}{\bibfnamefont{H.}~\bibnamefont{Barnum}},
  \bibinfo{author}{\bibfnamefont{C.}~\bibnamefont{Caves}},
  \bibinfo{author}{\bibfnamefont{C.}~\bibnamefont{Fuchs}},
  \bibinfo{author}{\bibfnamefont{R.}~\bibnamefont{Jozsa}}, \bibnamefont{and}
  \bibinfo{author}{\bibfnamefont{B.}~\bibnamefont{Schumacher}},
  \bibinfo{journal}{Phys. Rev. Lett.} \textbf{\bibinfo{volume}{76}},
  \bibinfo{pages}{2818} (\bibinfo{year}{1996}).

\bibitem[{\citenamefont{Kalev and Hen}(2008)}]{amir}
\bibinfo{author}{\bibfnamefont{A.}~\bibnamefont{Kalev}} \bibnamefont{and}
  \bibinfo{author}{\bibfnamefont{I.}~\bibnamefont{Hen}},
  \bibinfo{journal}{Phys. Rev. Lett.} \textbf{\bibinfo{volume}{100}},
  \bibinfo{pages}{210502} (\bibinfo{year}{2008}).

\end{thebibliography}
\end{document}